\documentclass[9pt,twocolumn,twoside]{osajnl}

\journal{ol} 

\setboolean{shortarticle}{true}
\usepackage{xcolor}
\usepackage{graphicx}

\title{Towards compact high-efficiency grating couplers for visible wavelength photonics}

\author[1,*]{Joe A. Smith}
\author[1]{Jorge Monroy-Ruz}
\author[1]{Pisu Jiang}%
\author[1]{John G. Rarity}
\author[1,2]{Krishna C. Balram}%

\affil[1]{Quantum Engineering Technology Labs and Department of Electrical and Electronic Engineering, University of Bristol, Bristol BS8 1UB, UK}
\affil[2]{krishna.coimbatorebalram@bristol.ac.uk}

\affil[*]{Corresponding author: j.smith@bristol.ac.uk}
\begin{abstract}
While grating couplers have become the de-facto standard for optical access to integrated silicon photonics platforms, their performance at visible wavelengths, in moderate index contrast platforms like silicon nitride, leaves significant room for improvement. In particular, the index contrast governs the diffraction efficiency per grating tooth and the resulting overall coupler length. In this work, we develop two approaches to address this problem: a dielectric grating that sums multiple optical modes to increase the overall output intensity, and an embedded metal grating that enhances the attainable refractive index contrast, and therefore reduces the on-chip footprint. We present experimental results that can be developed to realise compact efficient visible wavelength photonic interconnects, with a view towards cryogenic deployment for quantum photonics, where space is constrained and efficiency is critical.
\end{abstract}

\setboolean{displaycopyright}{false}

\begin{document}
\maketitle

The majority of atom (Rb - 780 nm) and atom-like (spin defects in diamond and silicon carbide) systems of interest for large-scale quantum technologies have their transition frequencies in the visible to near-infrared (IR) frequency range, above silicon's bandgap \cite{tiecke2014nanophotonic,atature2018material}. As they start to scale in complexity, developing robust methods to interface these quantum systems with light become critical to achieving system-level efficiency. Silicon nitride (SiN), a material with an established foundry infrastructure and well-developed high performance optical devices, is a natural candidate for building integrated quantum photonic devices around solid state defects. While SiN photonics is well-developed, from the perspective of quantum photonics, there are some key challenges that still need to be addressed that arise due to the low index contrast available in an SiN platform.

These problems are best illustrated by considering a grating coupler, a standard element in silicon photonics, but one whose performance in SiN leaves a lot to be desired. The low index contrast limits the achievable scattering efficiency per grating tooth and therefore requires long on-chip couplers that occupy a significant on-chip footprint \cite{mehta2017precise} and a limited diffraction efficiency, as obtained from standard foundry processes \cite{sarah}.

Efficient grating couplers have significant advantages for quantum photonics, with regards to interfacing emitters located at the chip centre and avoiding routing (and excess loss) to the chip edge.  Especially in cryogenic environments, as will be the case for most quantum photonics experiments, there are significant space constraints, which are relieved by being able to interface efficiently in the surface normal to the chip.

Photonics also offers non-dissipative, thermally isolated wiring for classical interconnects \cite{lecocq2021control,miller2017attojoule}. Electrical wiring is a bottleneck in cryogenic computing platforms as each wire contributes a thermal budget that offsets the available cooling power. This will undoubtedly rear its head for large scale quantum information processing \cite{scott2021timing} interfaced through thousands of wires to higher temperature electronics \cite{arute2019quantum}. Considering classical communication at visible light wavelengths ($\approx$ 637 nm), we can gain a 6$x$ reduction in grating footprint area compared to 1550 nm, whilst still harnessing established telecommunication protocols \cite{rajbhandari2017review,minotto2020visible}. The mode-matching adibatic taper length would also decrease by over 2$x$ \cite{milton1977mode}.  In systems with spatial and mechanical constraints, such as in cryostats, reducing the required fraction of chip area would free up valuable space for on-chip processing.

With this in mind, in this work, we develop two different coupling approaches. First we develop a compact free space outcoupler that can funnel emission into multiple modes to be collected by a high-NA objective, with a view towards improving the out-coupling efficiency of an emitter located in an SiN waveguide. In parallel, we use an embedded metal layer to increase the index contrast (and scattering efficiency) in the SiN platform. We aim to design sub-100 ${\mu}m^2$ footprint components with at least -10 dB coupling power, considering typical quantum emission efficiencies \cite{arcari2014near}, and to match comparable classical insertion losses\cite{rajbhandari2017review,pitwon2012firstlight}.

\begin{figure}[t!]
\includegraphics[width=\columnwidth]{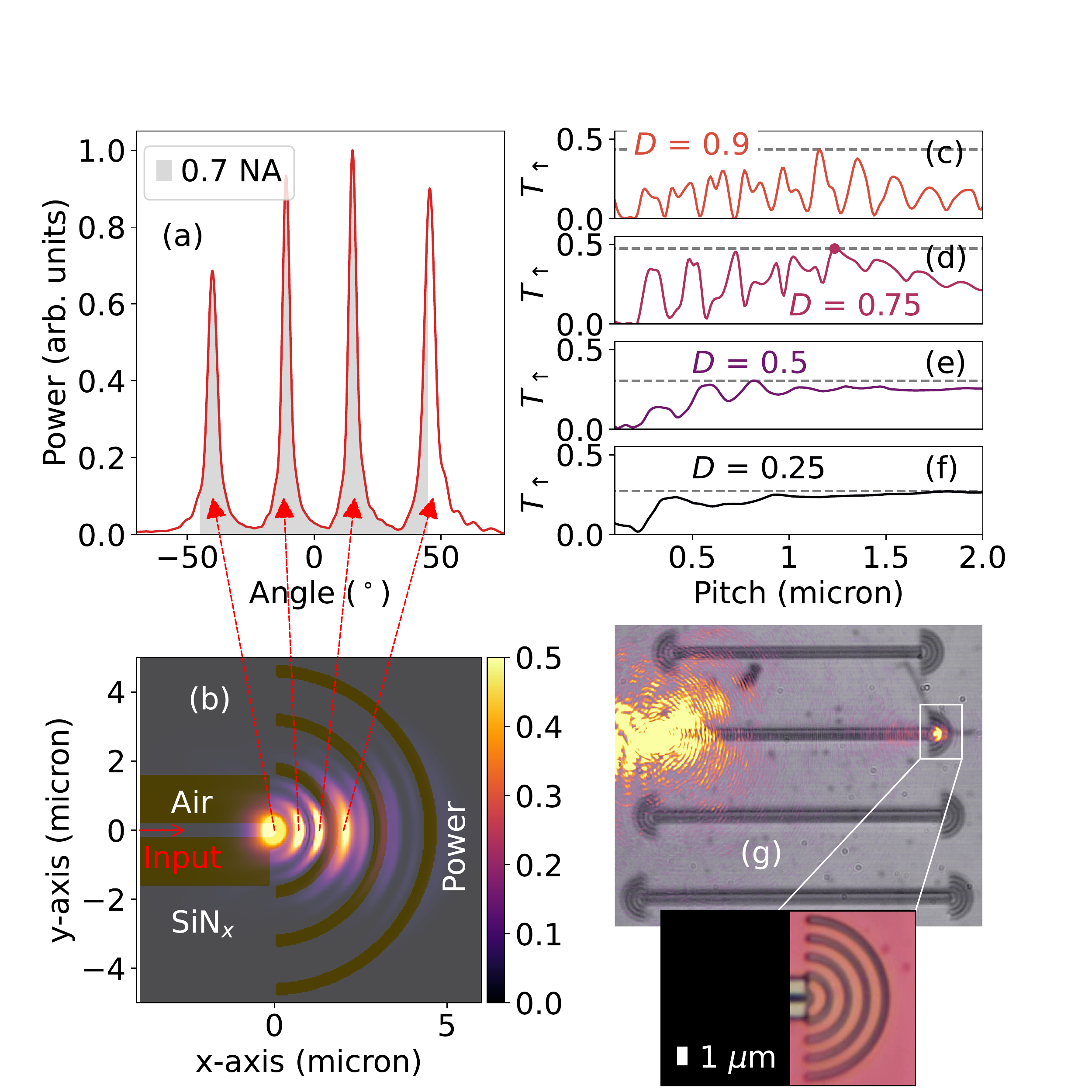}
\caption{(a) The far-field scattered power shows  42 \% of emission is accepted within NA = 0.7 (shaded in gray). (b) The 2D plot of power leaving the device, overlaid on the geometry, shows the multimode grating distributes into four angular lobes.  (c-f) As grating pitch increases, modes are excited leading to an increased collection $T_\uparrow$, maximised at duty $D \sim 0.75$, shown with purple dot. (g) In/out coupling and waveguiding of 637 nm laser light demonstrated through brightness of out-grating (right side). Scattered light off the in-grating (left side) saturates camera. Inset: microscope image of fabricated grating.
}\label{simulation-na}
\end{figure}

Compared to the moderately efficient small diffractive couplers reported in suspended gallium arsenide membranes \cite{arcari2014near,jiang2020suspended}, SiN on silica has two significant issues. Firstly, SiN has a considerably lower refractive index. Secondly, the silica bottom cladding means that light entering or exiting the grating will preferentially enter the substrate compared to a suspended platform, which is air clad on both sides. In addition, when imaging diffractive couplers in mechanically constrained systems, a long working distance objective reduces the available Numerical Aperture (NA). The 6.0 mm MY100X-806 Mitutoyo reduces the available NA = 0.7 or an acceptance angle of 45 degrees. Owing to this, light  will not be collected at large diffraction angles. Here we optimise a diffraction-limited grating coupler design within these limitations, using a 270 nm nitrogen-rich SiN film ($n$ = 1.89 at 637 nm) \cite{smith2019single}, with this thickness chosen to maximise refractive index contrast whilst maintaining single-mode operation.

Under these constraints, an optimal device operating at 637 nm is found by varying the design pitch and duty cycle of the grating $(\Lambda, D)$ and maximising the far-field projection of the emitted power (Poynting vector) that falls within the desired NA, $T_{NA}$, using a particle swarm algorithm in a numerical finite difference time domain solver (Lumerical FDTD) \cite{robinson2004particle}. The algorithm converges after 50 generations with the power collected in the top plane $T_\uparrow \sim 0.5$, with $T_{NA}$ = 0.48 for a design accepting angles up to 60$^\circ$ (0.9 NA) or  $T_{NA}$ = 0.42 accepting angles up to 45$^\circ$ (0.7 NA). The angular out-coupled power distribution of the 0.7 NA design is depicted in Fig \ref{simulation-na}(a). Here the design parameters are $(\Lambda, D)$ =  (1.41 $\mu$m, 0.77). It can be seen in this plot that there are four discrete lobes of emission and most of the emission is confined within the 0.7 NA acceptance angle of the objective (shaded in grey). These distinct lobes arise from the Bragg condition of the grating and match four emission regions in the horizontal cross-section in Fig \ref{simulation-na}(b). 

Considering that an air-suspended fully-etched structure would have $T_\uparrow$ = 0.5 by symmetry, our design is near optimal as the silica substrate reduces the top plane fraction.  The resolution required is 300 nm, within reach of standard lithography in a foundry-compatible platform. We note related work by Zhu \emph{et al.} on SiN grating couplers, to maximise 633 nm coupling into an NA = 0.65 objective \cite{zhu2017ultra}, however this relies on suspended SiN without a connecting substrate. Unlike this study, we vary the duty cycle $D$. The clear advantage of this is shown in Fig \ref{simulation-na}(c-f), as the collected power saturates for $D < 0.75$ as the pitch is increased. Further increase of $D$ above this has little value.

This device was fabricated using single step e-beam lithography with CSAR-62 resist to obtain a high ICP etch selectivity for the 270 nm SiN film \cite{thoms2014investigation}. A representative image of a fabricated device is shown inset in Fig \ref{simulation-na}(g). To characterise the grating, an input laser controlled by a steering mirror (PI - S334), translated through a $4f$ pair of lenses, sets the angle of entry into the objective and hence the excitation position at the sample. The sample is mounted on a stepper motor (PI MiCos) inside a cryostat. A beamsplitter on a flip mount allows white light illumination to align the laser on the grating. The collection is transmitted through a 90:10 beamsplitter and focussed onto a camera. Fig  \ref{simulation-na}(g) is taken with the white light image overlaid with the laser-only image showing alignment to the grating device. It evidences guiding of light through the waveguide structure with the right side coupler displaying a central spot with multiple crescent lobes of emission, as expected from the design. There is a large component of light scattered through the substrate from the in-coupler, shown in the broad extent of the input spot, which contributes to poor signal-to-noise. It was not possible to discriminate the out-coupled multi-modal signal above the surrounding noise floor, precluding quantitative analysis of this design. To reduce background scatter, a larger trench could be etched to separate the two grating couplers from the surrounding. However, there is a fundamental issue with this multi-mode grating, which is the cause of the scatter at the input beam. Primarily, although the device is a good out-coupler into several modes, by definition it is a poor in-coupler into a single mode waveguide from a focused Gaussian beam.

In this design, we have increased the amount of light coupled out of the device by collecting power summed from several different spatial modes, at a loss of modal purity, which also restricts the coupling efficiency obtainable into a single mode fibre. Owing to this, there is a limited overlap between a single Gaussian mode and the spatial profile of the light that couples out of the device, resulting in large scattering. At most a third of the emission would couple into a fibre, considering the intensity in a single lobe in Fig \ref{simulation-na}(a). This will be compounded in a chip with many grating couplers as the multi-directional emission and scattering will manifest as cross-talk between devices. Instead of considering only out-coupling, a single-moded design would equally be suitable for in-coupling. In order to achieve this, we shall establish an alternative type of grating  by considering bi-directional information transfer.

\begin{figure}[t!]
\includegraphics[width=\columnwidth]{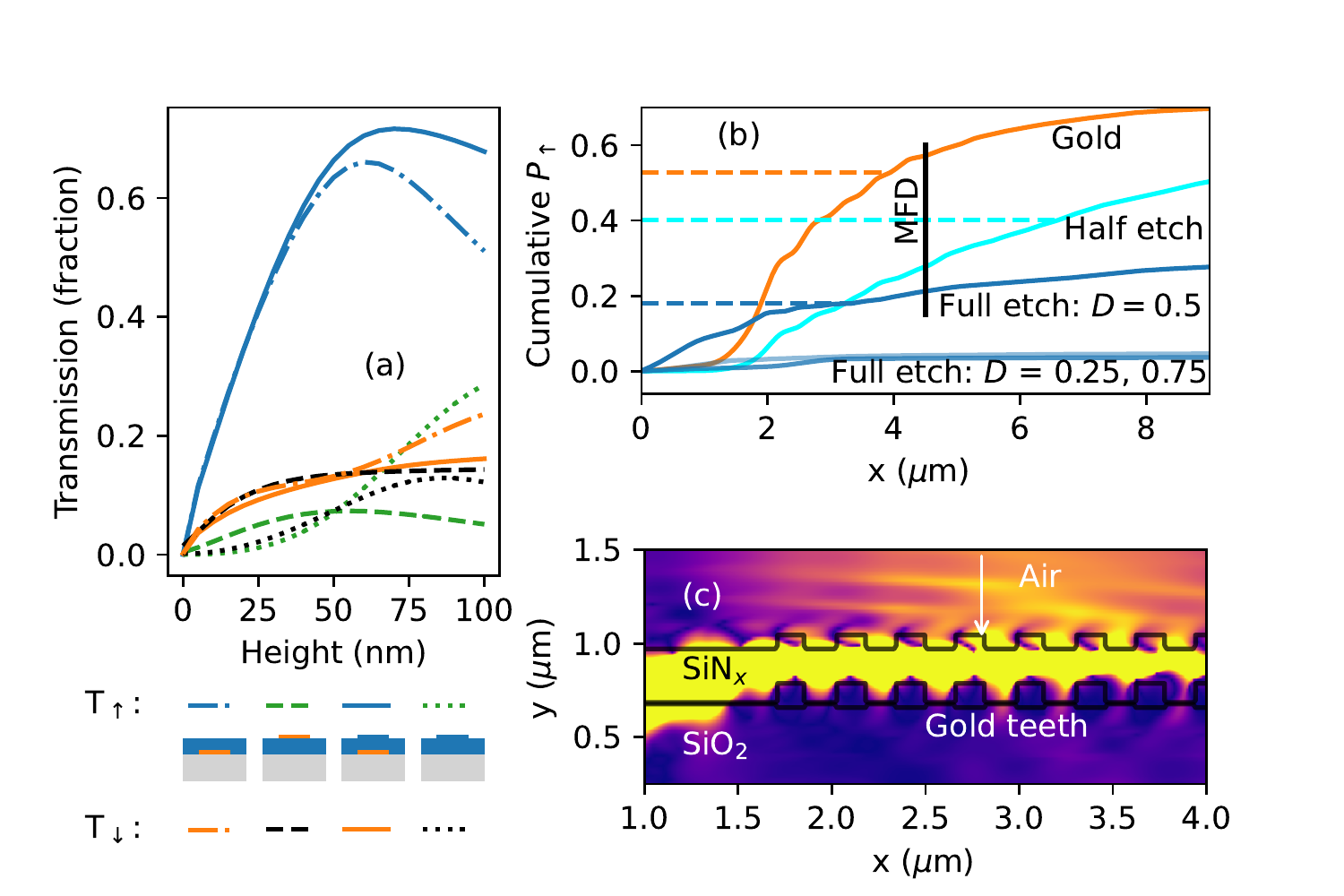}
\caption{(a) Transmission in the top ($T_\uparrow$) and bottom plane ($T_\downarrow$) as the grating height is varied for four symmetry-broken gratings. In the legend diagrams, SiN is shown in blue and gold in orange on a grey glass substrate. (b) FDTD simulation showing, as opposed to full and half etched dielectric gratings, the gold design results in a high proportion of emission leaving the device within the 630 nm MFD = 4.5 $\mu$m. (c) FDTD simulation showing input from simulated SM 630 fibre, with the majority of the input power coupled into the waveguide. 
}
\label{sim-metal}
\end{figure}

To realise efficient single-moded gratings that remain micron-sized, we must consider degrees of freedom that would increase the refractive index contrast. This is to ensure that most of the light leaves in the first few grating teeth of the device. As well as the need for a small grating length, by symmetry, emission in the top plane can not be greater than emission in the bottom plane in fully etched designs, even for an infinite grating length. One way to achieve a high-contrast broken symmetry design is to use a metallic layer in combination with a dielectric. Gold is highly reflective, with a low real component and high imaginary refractive index (here modelled as $n$  = 0.12 and $k$ = 3.33 at $\lambda$ = 637 nm \cite{johnson1972optical}).  We can break the symmetry of the device by patterning a gold grating before the SiN is deposited. This grating strongly perturbs the effective index of the SiN waveguide. Evaluating the effective index of 70 nm gold under 270 nm SiN as $n_{e}$ = 2.15, we see a 300 \%  increase in index contrast compared to a half-etched grating design ($n_e$ = 1.58:1.74).

In Fig \ref{sim-metal}(a), sweeping the thickness of the gold film, we find a maximal contrast between the transmission in the top plane ($T_\uparrow$) and the transmission in the substrate  ($T_\downarrow$) as $ T_\uparrow / T_\downarrow = 4 $ with a 70 nm-thick film. Simulations indicate that optimal contrast is obtained when this gold layer is deposited below the SiN (dash dotted line), rather than on top (dashed line), which instead increases emission into the substrate. We believe this is due to the large out of phase component of the metal which acts as a mirror, reflecting light where it would otherwise be transmitted. In the simulation, we model perturbations in the top of the grating due to the conformal nature of the SiN film deposition (solid line). These have negligible effect, thought due to the negligible change (<0.05\%) in the effective index of the thicker SiN structure. The taller grating structure has a slight symmetry breaking effect (dotted line), similar to partial etches, and has been demonstrated before \cite{marchetti2019coupling}.

Simulating the 70 nm buried gold device, we find that a pitch $\Lambda$ = 320 nm maximises outcoupling $T_\uparrow >0.5$. In contrast to the four lobes in the previous design, this emission leaves the grating design in a nearly mono-directional 20$^\circ$ from the substrate. The transition to this single angle emission also acts as an effective in-coupler of single mode Gaussian beams with matched spot sizes, by time reversal symmetry.

As this pitch requires tighter resolution tolerances, we opt for a linear grating for ease of process optimisation. We target a 5-$\mu$m-wide linear grating: this allows for smaller footprints than standard 12 $\mu$m telecommunications gratings, whilst coupling to a broad range of free space mode beams. The length of the grating is similar in dimension, with a low number of grating teeth. With this width, we find in Lumerical MODE that a linear taper can efficiently convert the waveguide mode with $<$ 1 dB loss in 15 $\mu$m compared to $\sim 100 \mu$m required for silicon photonics at 1550 nm.  Additionally, the device is wide enough to be tested with SM 630 fibre with a Mode Field Diameter (MFD) of 4.5 $\mu$m to decouple the free space optics performance from the devices under test. In  Fig \ref{sim-metal}(b), the gold grating emits over 75 \% of its emission within this MFD (dashed gold line). 

An alternative method to break symmetry is to employ a partial etch into the SiN layer. However, this results in a lower refractive index contrast to the fully etched grating. Therefore, a long grating is required for power in the waveguide mode to leave the device. As Fig \ref{sim-metal}(b) shows, a partially etched grating must be 6 $\mu$m long in order to emit 75 \% of its total efficiency ( shown by dashed line). 

Further to this, an attractive strategy for most use cases is to modulate the duty cycle of the grating to tailor emission to fit a Gaussian output mode  \cite{marchetti2017high,mehta2017precise}. Such phase-profiled apodised gratings in SiN are considerably larger in size as the initial duty cycle reduces the emitted power per grating tooth. In Fig \ref{sim-metal}(b), we see that either a reduced or increased $D$ results in a significant reduction in the power emitted over the same  length, compared to the $D$ = 0.5 grating. Therefore, duty cycle modulation is not beneficial for the device efficiency here, due to the limited number of grating teeth within the desired design length.

Following these results, we simulate the SM 630 fibre facet in FDTD, angled at 14 $^\circ$ in order to inject light into the device refracted through the air interface at 20 $^\circ$. In Fig \ref{sim-metal}(c), we can see that a high proportion of the power (> 0.5) is transmitted into the waveguide through the grating. The grating metal absorbs 8 \% of the incident power. With a cryostat cooling power of 1 W at 4K, this absorption sets a limit of $10^5$ devices for classical communication, assuming 100 $\mu$W of light per channel \cite{pitwon2012firstlight}. 

\begin{figure}[b!]
\includegraphics[width=\columnwidth]{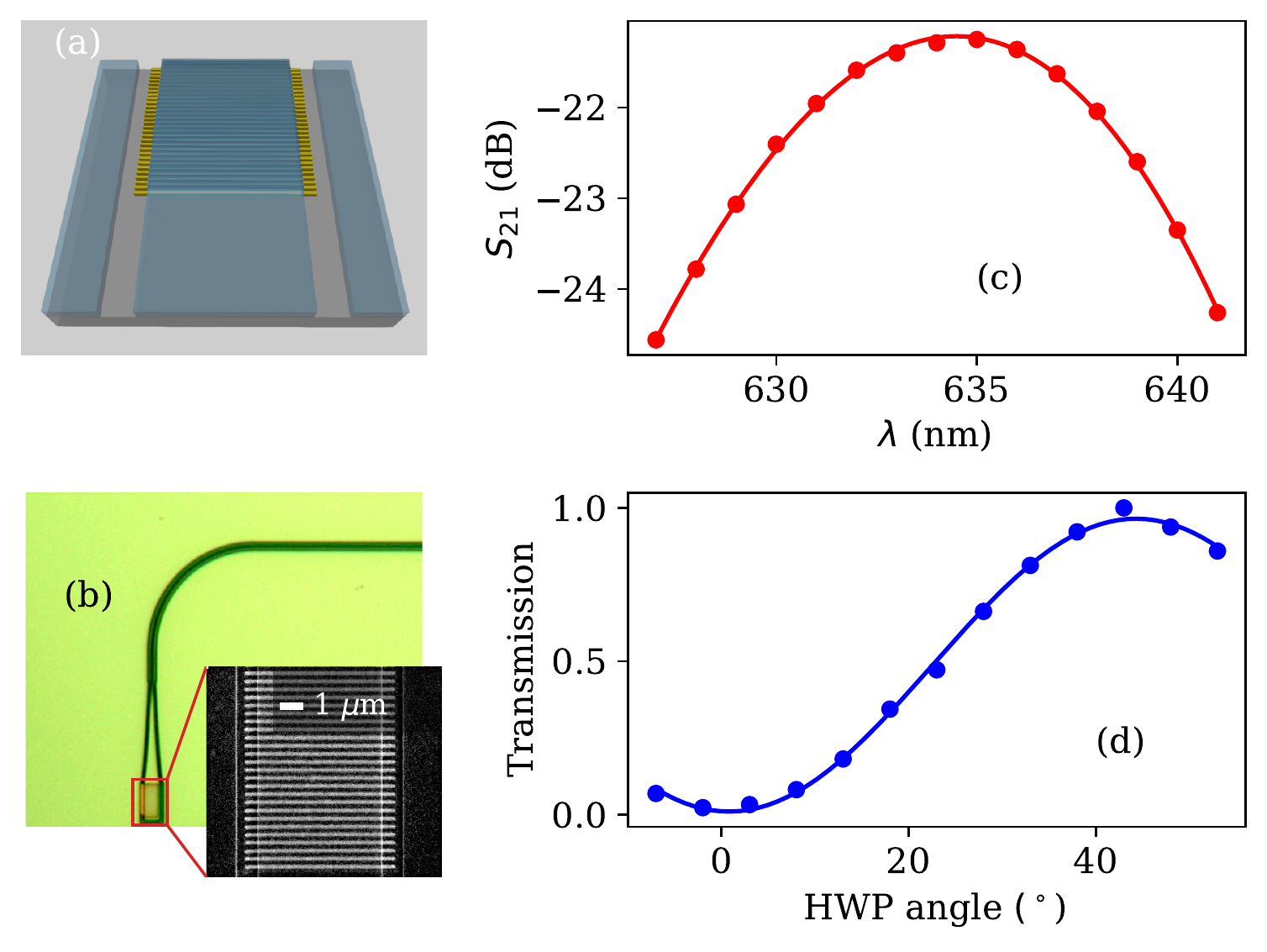}
\caption{(a) Schematic of gold grating. (b) One half of fabricated grating pair and connecting silicon nitride waveguide section. Inset: SEM of the 320 nm pitch gold grating. (c) Measured fibre-to-fibre device transmission has a $S_{21}$ = - 21 dB with a 3 dB bandwidth of 20 nm. (d) Changing the input polarisation with a half-waveplate, the device exhibits strong selectivity.}
\label{fab-metal}
\end{figure}

The grating coupler was fabricated through a lift-off process using PMMA. Gold is deposited on the patterned substrate using thermal evaporation and is lifted-off using a 10 second oxygen plasma de-scum and low power sonication in acetone. We avoid a chromium adhesion layer to reduce absorption losses and increase the scattering efficiency. Then, 270 nm of SiN is deposited on the grating using plasma enhanced chemical vapor deposition \cite{smith2019single}. The waveguide connecting the gratings is formed as in the previous device, with an inter-grating pitch of 250 um. The grating lines are extended on either side of the taper section such that the etch stage defining the grating can be overlaid with a $\pm$ 1 ${\mu}m$ tolerance. A schematic of the design is given in Fig \ref{fab-metal}(a). Fig \ref{fab-metal}(b) shows an image of the gold lift-off lines overlaid by the SiN waveguide. 

To test the device, we mount a V-groove fibre array containing 630 PM fibre (OZ Optics) with a polished facet at 14 $^\circ$ on a 6-axis fibre probing setup (Maple Leaf Photonics). The gratings are separated by 250 $\mu$m to match the V-groove spacing by a bend and linear waveguide section shown in  Fig \ref{fab-metal}(b). One channel is used to excite the grating and the other to readout the power from the adjacent grating using a Thorlabs PM100 power meter. Light from a supercontinuum source (NKT SuperK EVO EU-4) is controlled using an acousto-optic tunable filter (NKT SuperK SELECT) into the excitation channel. In  Fig \ref{fab-metal}(b), at its design wavelength of 637 nm, the best-performing device is observed to have an $S_{21}$ of -21 dB from fibre to fibre, indicating around -10 dB efficiency per coupler assuming lossless waveguiding. However, we observe some scattering of light along the waveguide in the camera indicating some waveguide loss.  It is known that roughness arising from liftoff has a significant effect in metallic nanostructures and can be replaced by gold etching in future devices \cite{greenwood2022smooth}. Residual roughness and e-beam overexposure also results in a higher effective $D$. We estimate the measured device has a $D$ = 0.7 and simulate this corresponds to a reduction in efficiency by almost half to $T = 0.28$. Film thickness variation is also significant with 10 nm error resulting in a further 20 \% loss. We would like to note here that although our measured devices are lower than the predicted efficiency, their performance is comparable to commercial SiN gratings at 1550 nm \cite{sarah}. 

The grating was designed for 637 nm, but our measurement peak is centered at 635 nm. The fibre array set angle is precise to 1$^\circ$ and this is likely a cause of this discrepancy. The measured device is highly broadband with a 3 dB bandwidth of 20 nm and so operates well at the design wavelength. In Fig \ref{fab-metal} (c), a half-waveplate (HWP) was inserted before the first grating to control the polarisation of the input light. From this measurement, coupling displays strong polarisation selectivity, as expected from a linear grating. This could have useful applications for reducing crosstalk by interleaving perpendicular devices.

We note that embedded metal gratings have been considered before. Wang \emph{et al.} simulated a gold grating below SiN although they did not consider the gold thickness \cite{wang2012embedded}. Lamy \emph{et al.} fabricated a gold grating for 1550 nm in the centre of titanium dioxide, where gold thickness is considered \cite{lamy2017broadband}.  Our demonstration experimentally realises a metal grating for visible wavelengths and specifically makes the connection that high-index contrast gratings are necessary for small single-moded designs. 

Moving forward, the designs presented here could have useful applications in high-density addressing of quantum emitters \cite{smith2019single, mehta2017precise}. SiN photonics can be deposited directly on metal when grown using low temperature PECVD, as presented here. Mature processes of this kind include the foundry platform developed by IMEC for integration with CMOS imagers \cite{malak2018monolithic}. Further work could look at a hybrid device to combine both structures into a single-mode in-coupler and multi-mode out-coupler centred around an integrated quantum emitter. We could extend our designs towards shorter wavelengths to potentially interface with trapped ions \cite{minotto2020visible}. Arrays should then be fabricated to measure channel crosstalk, as a step towards high-density photonic interconnects.

\section{Backmatter}

\begin{backmatter}
\bmsection{Funding} European Research Council (ERC-StG, SBS 3-5, 758843); Consejo Nacional de Ciencia y Tecnología (CONACyT); Engineering and Physical Sciences Research Council (EP/M024458/1, EP/N015126/1, EP/L015730/1); British Council (352345416).

\bmsection{Acknowledgments} We thank A. Murray, M. Cryan and D. Sahin for valuable suggestions. JAS and JGR were supported by British Council IL6 (352345416) and JGR's EPSRC Fellowship EP/M024458/1. JAS and JMR were supported by the EPSRC Quantum Engineering CDT  EP/L015730/1. JMR was supported by Consejo Nacional de Ciencia y Tecnologia (CONACyT). KCB acknowledges support from the ERC (ERC-StG, SBS 3-5, 758843). Device fabrication was performed using equipment acquired through EPSRC QuPIC EP/N015126/1.

\smallskip

\bmsection{Data availability} Data presented in this manuscript is available from the authors on request.  

\bmsection{Disclosures} The authors declare no conflicts of interest.

\end{backmatter}


\bigskip

\bibliography{bibl}


\end{document}